 \documentstyle[twocolumn,aps,prl,floats,epsf]{revtex}

\preprint{\bf PREPRINT}
\input psfig

\begin{document}
\wideabs{
% \columnsep0.1truecm
% \draft
%\preprint{ }
\title{Heat Capacity of $^3$He in Aerogel}
\author{Jizhong He, A.D. Corwin, J.M. Parpia, J.D. Reppy}
\address{Laboratory of Atomic and Solid State Physics and the CCMR, Cornell University, Ithaca, NY 14853-2501}

\date{\today}

\maketitle

\begin{abstract}

The heat capacity of pure $^3$He in low density aerogel is measured at
22.5 bar.  The superfluid response is simultaneously monitored with a
torsional oscillator.  A slightly rounded heat capacity peak, 65
$\mu$K in width, is observed at the $^3$He-aerogel superfluid
transition, $T_{ca}$.  Subtracting the bulk $^3$He contribution, the
heat capacity shows a Fermi-liquid form above $T_{ca}$.  The heat
capacity attributed to superfluid within the aerogel can be fit with a
rounded BCS form, and accounts for 0.30 of the non-bulk fluid in the
aerogel, indicating a substantial reduction in the superfluid order
parameter consistent with earlier superfluid density measurements.

\end{abstract}
}

The 1995 discovery of the superfluid transition of $^3$He in aerogel
glass\cite{porto95}\cite{sprague95}, has aroused wide interest.  The
$^3$He-aerogel system presents a unique opportunity for study of the
influence of quenched impurities on the well-understood system of pure
superfluid $^3$He.  Earlier studies employing other porous media such
as sintered silver or packed powders \cite{kojima}\cite{parpia92} were
dominated by surface scattering and finite size effects\cite{ambeg}.
Aerogel consists of a fractal structure of relatively uniform silica
strands 30 \AA\ in diameter.  In the case of 97.6\% open aerogel,
employed in the present experiment, scattering due to the silica
strands, though strong enough to introduce a significant quasiparticle
density at T=0, is sufficiently weak so that superfluidity is not as
strongly suppressed as it would be the case of more dense aerogel
systems or packed powders.  While previous measurements exposed the
onset of superfluidity using torsional pendulem
\cite{porto95}\cite{matsumoto}, NMR \cite{sprague95} \cite{sprague96}
\cite{alles99} \cite{barker00} \cite{bunkov00prl} and acoustic
techniques\cite{acoustic} \cite{halperin}, they could not distinguish
between a percolation transition, where the flow and the specific heat
anomaly are distinct, and a suppressed superfluid
transition\cite{stroud}.

% Figure 1
\begin{figure}
\centerline{\psfig{figure=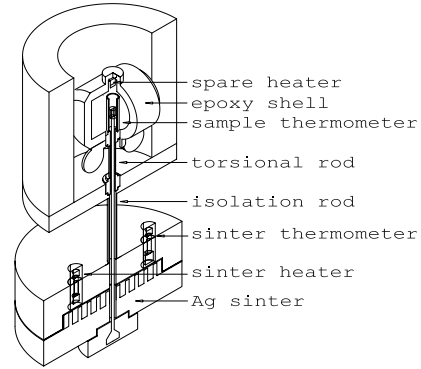,width=2.2truein}}
\caption{A cross-section of the experimental cell is shown.}
\label{cell}
\end{figure}

In the work presented here, we examined the $^3$He B phase superfluid
state in aerogel through heat capacity measurements at a pressure of
22.5 bar\cite{alles99,barker00}.  Our chief result is the observation
of a feature in the heat capacity coincident with the onset of
superflow in the aerogel-$^3$He system.  This observation confirms
that superflow onset corresponds to a true phase transition rather
than to a possible dynamic effect associated with a percolation
transition involving distributed superfluid regions.  The analysis
reveals a normal fermi liquid like contribution to the heat capacity
down to the lowest temperatures, corresponding to the existence of a
non-zero normal fraction down to absolute zero.  In earlier work
\cite{qfs00}, we reported on measurements of the heat capacity
employing a drift technique.  Those measurements gave an indication of
an anomaly at the aerogel superfluid transition; however, a definitive
resolution of this peak has required the improvements of the present
experiment.

In Fig. 1, we show a cut-away diagram of the latest version of our
heat capacity apparatus.  The cell containing the $^3$He-aerogel
sample also forms the inertial head of a torsional oscillator.  We are
thus able to correlate features in the heat capacity with the dynamic
superfluid response.  The liquid $^3$He within the apparatus is cooled
by a silver sinter pad, which is thermally clamped to a PrNi$_5$
nuclear cooling stage.  There are two lanthanum diluted cerium
magnesium nitrate (LCMN) thermometers; one inside the sample cell and
the other located near the silver sinter. These are operated in a dc
mode similar to that of the SQUID based thermometer developed by Lipa
and Chui \cite{lipa}.  Thermometer calibration is provided by
comparison to a melting curve thermometer mounted on the nuclear
cooling stage\cite{greywall86}. Thermal contact between the heat
capacity cell and the cooling stage is established via the $^3$He
contained in the hollow torsion rod that connects the heat capacity
cell to the stage.  The time constant associated with this thermal
contact path is designed to be at least an order of magnitude longer
than the thermal relaxation time within the heat capacity cell itself.
Heat pulses are applied to the sample by a non-inductive resistive
heater wound around the aerogel sample.  Care is taken that the heater
is not close to the LCMN thermometer.  In addition, a second spare
heater, shown in Fig. 1, was placed in the cell.

% Figure 2
\begin{figure}
\centerline{\psfig{figure=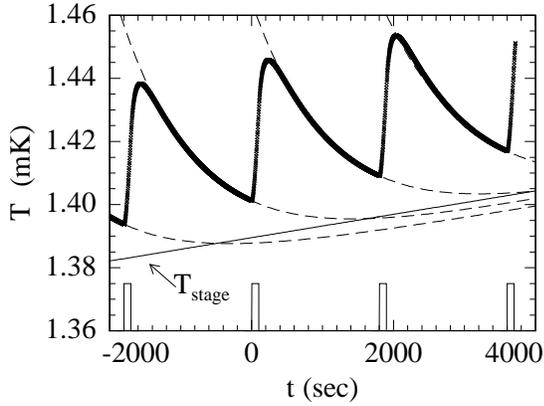,width=3.1truein}}
\caption{The thermal response of the cell thermometer to a series of
four heat pulses (depicted at the bottom of the figure) is shown as a
function of time.  The solid line indicates the steady drift of the
stage temperature and the dashed curves are fits of the exponential
decay function (see text) to the relaxation portion of the cell's
thermal response following a heat pulse.}
\label{3pulses}
\end{figure}

Our measurement technique is illustrated in Fig. 2.  After a
demagnetization to our lowest temperatures, the cell warms slowly over
several days.  During this warming interval, a sequence of heat pulses
is applied to the $^3$He within the cell and the heat capacity is
determined as a function of temperature.  Representative data are
shown in Fig. 2, illustrating a series of heat pulses and the
subsequent thermal relaxation of the sample cell temperature,
$T_c(t)$.  In this example, the stage temperature $T_s(t)$, is
increasing at a steady rate of 15.6 $\mu$K/hr and identical heat
pulses are repeated at 1800 second intervals.  Following each heat
pulse, the cell temperature rises rapidly.  Initially the temperature
distribution within the cell is nonuniform.  The internal equilibrium
time is relatively short, so that over time the thermal distribution
within the cell becomes nearly uniform.  The cell temperature exhibits
an exponential relaxation of the form, $T_c(t) =
T_{0}e^{-t/\tau}+T_{s}(t)-\alpha \tau$ , where $\alpha$ is the warming
rate of the stage and $\tau$ is the relaxation time constant.  Fits of
this function to the relaxation portion of the thermal pulses are
shown in the figure.

The temperature excursion during each heat pulse is relatively small,
on the order of 40 to 50 $\mu$K.  Therefore, in our analysis we take
the heat capacity, C, and thermal conductivity, $\kappa$, to be
constant over the period of each relaxation.  We proceed by
calculating the net energy flow, Q, into or out of the cell during a
time interval, $t_1$ to $t_2$, restricting these times to the
exponential portion of the thermal relaxation of the cell.  During
this period the LCMN thermometer gives a reliable value for the
temperature throughout the cell. The energy flow from the cell, due to
thermal conduction down the torsion rod to the stage, is obtained by
numerical integration of the quantity $\kappa (T_{c}(t) - T_{s}(t))$,
over the chosen time interval.  If the time interval contains the heat
pulse, we include the energy, $Q_{in}$, contributed by the heater.  A
possible choice for the first time interval would be to take $t_1 =
0$, the time just before the heat pulse, and $t_2 = t_f$, the time at
the end of the relaxation period just before the next heat pulse. The
second interval, which will not include a heat pulse, might start at
time $t$ (chosen in the exponential decay region) and ends at $t_f$ as
for the first interval. Then we write two independent equations
containing the heat capacity and the thermal conductivity.

\begin{equation}
C(T_c(t_f)-T_c(0)) = Q_{in} - \kappa \int_0^{t_{f}} ( T_c(t) - T_{s}(t))dt
\end{equation}

\begin{equation}
C(T_c(t_f)-T_c(t)) = - \kappa \int_t^{t_{f}} ( T_c(t) - T_{s}(t))dt
\end{equation}

 These equations are then solved for C and $\kappa$ with $A$ defined
as $\int_0^t (T_c(t) - T_s(t))dt$ and $A_f$ defined as $\int_0^{t_{f}}
(T_c(t) - T_s(t))dt$.

\begin{equation}
C=\frac {(A_{f}-A)Q} {A_{f}(T_c(t)-T_c(0))-A(T_c(t_f)-T_c(0))}
\end{equation}

\begin{equation}
\kappa = \frac{(T_c(t)-T_c(t_f))Q}{A_{f}(T_c(t)-T_c(0))-A(T_c(t_f)-T_c(0))}
\end{equation}

 One advantage of this approach is that one can vary $t$, and check
that the calculated values for C and $\kappa$ are independent of $t$.
The thermal conductivity obtained from the above analysis is found to
be a smoothly varying function of temperature except at the bulk
superfluid transition where the conductivity drops by over a factor of
two as the temperature rises through the transition temperature.  A
more detailed discussion of the data analysis and the cell
construction will be found elsewhere \cite{qfs01} \cite{corwinthesis}.

Throughout the discussion that follows the bulk contribution is taken
to arise from an equivalent amount of superfluid in the absence of the
aerogel.

% Figure 3
\begin{figure}
\centerline{\psfig{figure=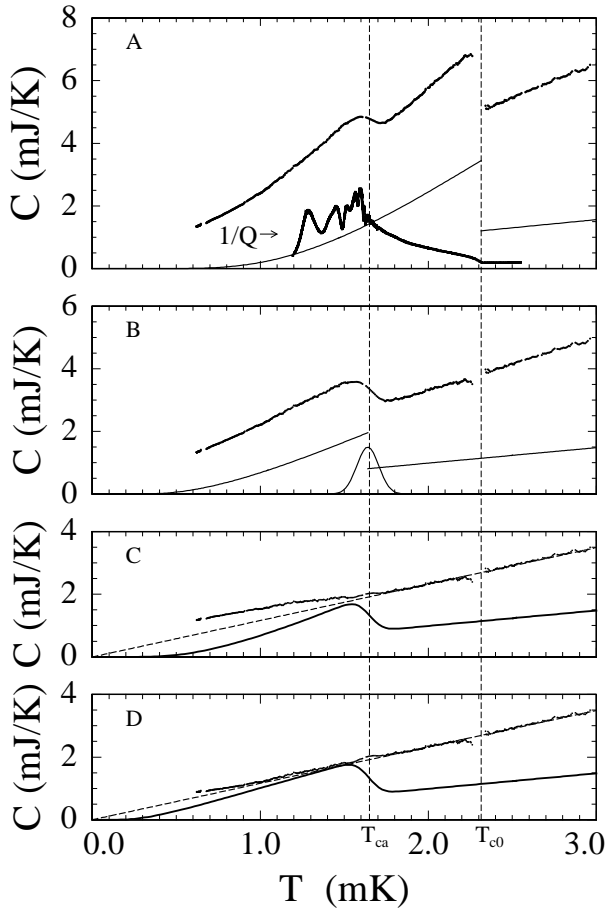,width=3.6truein}}
\caption{Panel A shows the total measured heat capacity for the
$^3$He-aerogel sample at a pressure of 22.5 bar, the calculated bulk
$^3$He contribution, and the dissipation, $Q^{-1}$, of the torsional
oscillator.  Panel B shows the heat capacity after the bulk
contribution is subtracted, along with the BCS heat capacity form and
the gaussian critical temperature distribution used for the
convolution.  Panels C and D show the residual heat capacity after the
convolved BCS heat capacity has been subtracted along with the
convolved heat capacity, with $\delta_{sc}$ = 1.762 for panel C, and
$\delta_{sc}$ = 0.88 for panel D.  The dashed lines are linear fits to
the normal fermi liquid remaining above $T_{ca}$.}
\label{fig3}
\end{figure}

In Fig. 3A, we have plotted the total heat capacity determined for
temperatures between 0.6 to 3 mK.  We also show the $Q^{-1}$ for the
torsional oscillator.  There are two conspicuous features in the heat
capacity data.  The first is the heat capacity anomaly associated with
the superfluid transition in the $^3$He-aerogel sample, which
coincides with the superfluid aerogel transition, $T_{ca}$, as marked
by the torsional oscillator.  The second feature is the sharp jump in
the heat capacity due to bulk $^3$He within the cell.  Above the bulk
transition, $T_{c0}$, a linear temperature dependence is seen, as
expected for a normal Fermi liquid.  We also note that in addition to
the linear term in this region there appears to be a small additional
constant contribution to the heat capacity ($\approx10\mu$J/K).  Golov
and Pobell suggest that this contribution may arise from ordering in
the amorphous solid $^3$He layer \cite{golov96}.

Since our main interest is in the heat capacity associated with the
$^3$He-aerogel superfluid transition, we shall subtract the bulk
$^3$He heat capacity contribution.  For this purpose we use the data
of Greywall \cite{greywall86} and multiply by a suitable factor so as
to produce a smooth continuation of the remaining heat capacity data
across the temperature of the bulk transition.  The magnitude of the
subtracted heat capacity indicates that a fraction (0.295) of the
sample has the heat capacity of bulk $^3$He.  This fraction is more
than we can reasonably account for in terms of cracks and small
volumes outside of the aerogel itself.  Therefore we conclude that our
sample of aerogel must contain a number of macroscopic pores.

% Figure 4
%\begin{figure}
%\centerline{\psfig{figure=capnobulk.ps,width=3.1truein}}
%\caption{The heat capacity remaining after the subtraction of
%the bulk contribution is shown as a function of temperature.  The transition temperatures, $T_{ca}$ and $T_{c0}$, as measured in the oscillator response are indicated by the arrows.} 
%\label{capnobulk}
%\end{figure}

The heat capacity data remaining after the subtraction of the bulk are
shown in Fig. 3B.  The heat capacity anomaly associated with the
aerogel transition shows rounding but is clearly separated from the
bulk transition. The linear Fermi liquid region extends smoothly from
temperatures above the bulk transition right to the transition in
aerogel.  In the region between the bulk transition and the aerogel
$T_c$ we find no indications of the anomalous behavior suggested by
recent Grenoble NMR experiments \cite{bunkov00prl}.

The $^3$He in aerogel effective mass, $m_a^*$, can be determined from
the slope of the heat capacity in the Fermi-liquid region
\cite{greywall86} \cite{wheatley75} after the fraction of bulk $^3$He
has been subtracted.  From the slope of the data of Fig. 3B above
$T_{ca}$, we find $m_a^*/m = 6.5$, approximately 30\% larger than the
ratio for bulk superfluid $^3$He at this pressure.  The apparent
enhancement of $m^*$ cannot be ascribed to a miscounting of the number
of $^3$He atoms in the cell, and we believe that it must reflect a
change in the excitations induced in the normal state by the presence
of aerogel.

The simplest theoretical approach to the problem of superfluid $^3$He
in aerogel is that of the Homogeneous Scattering Model (HSM)
\cite{thuneberg}, which treats the aerogel as homogeneous collection
of scattering centers.  The HSM predicts a reduction of the transition
temperature comparable to that seen in our experiments, as well as
suppression of the order parameter.

In order to assess the fraction of fluid contributing to the aerogel
superfluid transition, we have fit the data to an interpolation of the
BCS specific heat (Eq. 5)\cite{bcs}.  We take the weak coupling value
of $\Delta C/C = 1.42$, a transition temperature, $T_{ca}$, of 1.65
mK, $t=T/T_{ca}$, the normal fermi liquid heat capacity contribution,
$C_N(T_{ca^+})$, and the weak coupling value of $\delta_{sc} =
\frac{\Delta(0)}{k_B T_{ca}} \approx 1.764$.

\begin{eqnarray}
C_{s}(T)=C_N(T_{ca}^+)\Bigg(\frac{3}{\pi^2}\frac{\delta^2_{sc}}{t}\sqrt{2 \pi \frac{\delta_{sc}}{t}} (1-t^2) e^{-\frac{\delta_{sc}}{t}}+ \nonumber \\
\sqrt{t}\left(1+\frac{\Delta C}{C}\right) e^{-\delta_{sc}(\frac{1}{t}-1)}\Bigg),
\end{eqnarray}

The BCS specific heat is shown in figure 3B.  Since the peak in the
measured heat capacity is somewhat rounded, we fit it with a
convolution of the BCS form with a gaussian distribution of critical
temperatures, and display the residual heat capacity in figure 3C,
along with the convolved BCS heat capacity.  The best fit is from a
gaussian of width $65\mu K$, and the gaussian is displayed in figure
3B.  The dashed line in figure 3C is a linear fit to the normal fermi
liquid remaining above $T_{ca}$.  The residual heat capacity below
$T_{ca}$ rises above this line, suggesting other contributions to the
low temperature heat capacity.  This might be expected as a
consequence of the magnetic ordering of the solid $^3$He substrate
atoms, but this contribution is expected to be on the order of the
temperature independent constant at our lowest temperature
\cite{greywall88}.  Such an ordering of the solid layer is a feature
observed in the earlier NMR aerogel experiments \cite{sprague95}
\cite{sprague96} \cite{alles99} \cite{barker00}.  As has been
demonstrated by the NMR \cite{sprague96} \cite{barker00}, the
influence of the magnetic solid $^3$He layer can be removed by the
addition of a few layers of $^4$He on the substrate.  This will be an
interesting direction to pursue in future aerogel heat capacity
experiments.  

The implied normal fermi liquid fraction is unchanged by choice of
$\delta_{sc}$, and remains at 0.30.  If we apply the results of Lawes
and Parpia for the reduction of the energy gap for our $T_{ca}/T_{c0}$
of 0.71 \cite{lawes02}, we find $\delta_{sc} = 0.88$.  The residual
heat capacity computed for this value is shown in figure 3D, and shows
much less deviation from normal fermi liquid heat capacity (drawn as a
dotted line) than that of weak coupling energy gap value.  At lowest
temperatures, the residual rises above the normal fermi liquid, still
suggesting an extra heat capacity contribution, though much less than
that implied from figure 3C.

Somewhat surprisingly, we find that the fraction of the fluid that
appears to participate in the aerogel superfluid state is only a
fraction, 0.30, of the total.  Expressed in terms of the order
parameter, one would conclude that there has been a reduction by a
factor of over two under the conditions of our experiment.  This
result is in keeping with the order parameter reductions observed in
the Cornell torsional oscillator experiments
\cite{porto95}\cite{lawes02}, and in agreement with the gapless
superfluid model of Sharma and Sauls \cite{sharma01}.

% Figure 5
%\begin{figure}
%\centerline{\psfig{figure=nfl.eps,width=3.1truein}}
%\caption{The normal Fermi liquid in the aerogel seen after subtracting
%a BCS-like superfluid transition at $T_{ca}$ (shown as the solid
%line).  Note the upturn at lowest temperatures, suggesting magnetic
%ordering.}
%\label{nfl}
%\end{figure}

In conclusion, we find that the behavior of $^3$He confined to aerogel
shows normal Fermi liquid characteristics above the superfluid
transitions.  The enhancement of the effective mass along with the low
temperature excitations contribution in this region can be further
studied as a function of pressure in future research.  While a
sizeable fraction of fluid in the cell exhibits bulk-like
characteristics, a comparable fraction displays a BCS-like heat
capacity centered at the reduced $T_c$, in conformity with
expectations of a suppressed phase transition.  Thus there appears to
be a significant contribution (30\%) of ``bulk like'' gapped
excitations, and evidence for a gapless contribution to the specific
heat in conformity to observations in torsional pendulum measurements
\cite{porto95}.

% \acknowledgments

The authors are especially grateful to Prof. Norbert Mulders of the
University of Delaware who supplied the aerogel sample used in this
experiment.  We also wish to recognize the contributions of
A.L. Woodcraft and G.M. Zassenhaus to early development of this
experiment, and also acknowledge valuable discussions with Prof's
M.H.W. Chan and T.L. Ho, and Dr. E.N. Smith. One of us, J.D.R. would like
to acknowledge the Institute for Solid State Physics, University of
Tokyo, for hospitality during the period this paper was prepared.  The
research has been funded by the NSF under grants DMR-0071630 and
DMR-9971124.

\end{document}